\documentclass[11pt]{article}
\usepackage[utf8]{inputenc}
\usepackage{amsmath,amsxtra,amssymb,latexsym,amscd,amsthm,pb-diagram,fancyhdr,euscript}
\usepackage{amsthm}
\usepackage{indentfirst}
\usepackage{epsfig}
\usepackage{mathrsfs}
\usepackage{slashed}
\usepackage{hyperref}
\hypersetup{
   colorlinks,
    citecolor=blue,
    filecolor=black,
    linkcolor=red,
    urlcolor=magenta
}
\hypersetup{linktocpage}

\newcommand{\s}{\mathrm{S}}

\newcommand{\R}{\mathbb{R}} \newcommand{\N}{\mathbb{N}}
\renewcommand{\S}{\mathbb{S}}

\renewcommand{\d}{\mathrm{d}}

\newtheorem{theorem}{Theorem}[section]
\newtheorem{proposition}{Proposition}[section]

\newtheorem{lemma}{Lemma}[section]
\newtheorem{remark}{Remark}[section]
\begin{document}
\mbox{} \thispagestyle{empty}

\begin{center}
\bf{\Huge Geometry of Vaidya spacetimes}

\vspace{0.1in}

{Armand COUDRAY$^\mathrm{1}$ \& Jean-Philippe NICOLAS\footnote{LMBA, UMR CNRS 6205, Department of Mathematics, University of Brest, 6 avenue Victor Le Gorgeu, 29200 Brest, France.
Emails~: Armand.Coudray@univ-brest.fr, Jean-Philippe.Nicolas@univ-brest.fr.}}
\end{center}

{\bf Abstract.} We investigate the geometrical structure of Vaidya's spacetime in the case of a white hole with decreasing mass, stabilising to a black hole in finite or infinite time or evaporating completely. Our approach relies on a detailed analysis of the ordinary differential equation describing the incoming principal null geodesics, among which are the generators of the past horizon. We devote special attention to the case of a complete evaporation in infinite time and establish the existence of an asymptotic light-like singularity of the conformal curvature, touching both the past space-like singularity and future time-like infinity. This singularity is present independently of the decay rate of the mass. We derive an explicit formula that relates directly the strength of this null singularity to the asymptotic behaviour of the mass function.

\vspace{0.1in}


\vspace{0.1in}


\tableofcontents

\section{Introduction}

In 1959, P.C. Vaidya published a paper \cite{Va1943} in which he was solving a long standing open problem in general relativity: finding a modification of the Schwarzschild metric in order to allow for a radiating mass. His original derivation of the metric was based on Schwarzschild coordinates. Ten years later, in \cite{Va1953}, he observed that using instead what is now known as Eddington-Finkelstein coordinates would simplify the construction a great deal. Vaidya's metric is a solution to the Einstein equations with matter in the form of null dust and it has since been the object of numerous studies; see the book by J.B. Griffiths and J. Podolsky \cite{GriPo}, section 9.5, for a very clear presentation of the metric and an excellent account of the history of these investigations. Many of these works aimed at gluing some part of Vaidya's spacetime with other exact spacetimes like Schwarzschild and Minkowski, in order to construct models for evaporating or collapsing black holes, see for instance W.A. Hiscock \cite{Hi}. A different approach consists in studying the maximal extension (not necessarily analytical) and the matching of the Vaidya exterior with some interior metric. This has been done under explicit assumptions on the manner in which the mass varies with time, for instance in \cite{BeDoEro2016}. The general problem was first tackled by W. Israel \cite{Isra} and more recently by F. Fayos, M.M. Martin-Prats and M.M. Senovilla \cite{FaMaSe} and F. Fayos and R. Torres \cite{FaTo}, this last work studying the possibility of a singularity-free gravitational collapse.

In this paper, we study the geometry of Vaidya's spacetime itself, without extension, in as much generality as possible. We treat the case of a radiating white hole that emits null dust and as a result sees its mass decrease. The case of a black hole whose mass increases due to incoming null dust is obtained by merely reversing the arrow of time. We make only minimal natural assumptions on the behaviour of the mass function.
In the existing literature, many of the precise behaviours of null geodesics are studied numerically. In contrast, the goal of the present paper is to give a mathematical derivation of the geometrical features of Vaidya's spacetime, by analysing precisely the ordinary differential equation describing the incoming principal null geodesics, among which are the generators of the past horizon. This equation is well-known and appears explicitly in I. Booth and J. Martin \cite{BoMa}, in which a notion of distance between the event horizon and an apparent horizon is introduced. This is done for an evaporating white hole that, asymptotically or in finite retarded time, stabilises to a black hole or evaporates completely. We devote special attention to the case of a complete evaporation in infinite time (the case where the mass function vanishes in finite time has been studied in details in \cite{FaTo}, with a mass function that is  possibly non differentiable at its vanishing point), for which the spacetime is already maximally extended and, we prove the existence of a light-like singularity of the Weyl tensor, touching both the past spacelike singularity and future timelike infinity\footnote{Note that F. Fayos and R. Torres \cite{FaTo} also exhibit a null singularity but under slightly different assumptions. The evaporation ends in finite time, therefore the singularity is not asymptotic but well present in the spacetime. Moreover the evaporation is allowed to end brutally: the mass vanishes in finite retarded time with a possibly non-zero slope, in which case it is non-differentiable at its vanishing point.}. This classical evaporation in infinite time may perhaps be seen as a classical analogue of the final pop in black hole quantum evaporation, with the emission of a gravitational wave propagating along a light-cone.

The article is structured as follows. In Section \ref{GeomVaidya}, we recall the construction of Vaidya's spacetime as a simple modification of the Schwarzschild metric expressed in retarded Eddington-Finkelstein coordinates ($u=t-r_*,r,\theta,\varphi$) and we recall the essential features of the geometry: Petrov-type D, principal null directions, expression of the Christoffel symbols, Weyl and Ricci tensors in the Newman-Penrose formalism, scalar curvature and also the curvature scalar (also referred to as the Kretschmann scalar). We also present our natural assumptions on the mass function, which are essentially that it is smooth, decreases and admits finite limits as the retarded times tends to $\pm \infty$. Finally, we derive the ordinary differential equation that allows to locate the past event horizon and gives the complete congruence of incoming principal null geodesics. Section \ref{Incoming} presents the analysis of the properties of the solutions to this ODE and the construction of an optical function such that its level hypersurfaces are spanned by the incoming principal null geodesics. This is the generalisation of the function $v=t+r_*$ in the Schwarzschild case. In Section \ref{Evaporation}, we give further properties of the principal null geodesics in the case of a complete evaporation in infinite time and we prove that they all end up in the future at a light-like conformal curvature singularity. This singularity is present independently of the speed at which the mass function approaches zero in the future, however its strength seems to be directly related to the decay rate of the mass.  We also construct families of timelike curves that end up either at the null singularity or at timelike infinity depending of their ``mass'', i.e. the rate of their proper time as measured with respect to the retarded time. The section ends with Penrose diagrams of Vaidya's spacetime in the case of a complete evaporation in infinite time, showing the various congruences.

All formal calculations of connection coefficients and curvature tensors have been done using Sage Manifolds \cite{Go}.

{\bf Notations.} Throughout the paper, we use the formalisms of abstract indices, Newman-Penrose and $2$-component spinors.

\section{Vaidya's spacetime, connection, curvature} \label{GeomVaidya}

The Vaidya metric can be constructed as follows. We start with the Schwarzschild metric
\[ g = \left( 1-\frac{2M}{r} \right) \d t^2 - \left( 1-\frac{2M}{r} \right)^{-1} \d r^2 - r^2 \d \omega^2 \, ,~ \d \omega^2 = \d \theta^2 + \sin^2 \theta \, \d \varphi^2 \, .\]
We express it in outgoing Eddington-Finkelstein coordinates $(u,r,\theta,\varphi)$, where $u=t-r_*$ and $r_* = r+2M \log (r-2M)$:
\[ g= \left( 1-\frac{2M}{r} \right)  \d u^2 + 2 \d u \d r - r^2 \d \omega^2\, .\]
Vaidya's metric is then obtained simply by allowing the mass $M$ to depend on $u$:
\begin{equation} \label{VaidyaMetric}
g = \left( 1-\frac{2M(u)}{r} \right)  \d u^2 + 2 \d u \d r - r^2 \d \omega^2 \, .
\end{equation}
Throughout the paper, we assume that
\begin{equation} \label{MSmooth}
M \mbox{ is a smooth function of the retarded time } u
\end{equation}
and we denote
\[ \dot{M}(u) := \frac{\d M}{\d u} \, .\]
The non-zero Christoffel symbols for \eqref{VaidyaMetric} are
\begin{gather*}
\Gamma_{ 0 \, 0 }^{ 0  }  =  -\frac{M(u)}{r^{2}} \, ,~ \Gamma_{ 2 \, {2} }^{ \, 0}  =  r \, ,~ \Gamma_{ 3 \, {3} }^{ \, 0}  =  r \sin^2 \theta \, , \\
\Gamma_{ 0 \, 0 }^{ 1  }  =  -\frac{r^{2} \dot{M}(u) - r M(u) + 2 \, M(u)^{2}}{r^{3}} \, ,~ \Gamma_{ 0 \, 1 }^{ 1  }  =  \frac{M(u)}{r^{2}} \, , \\
\Gamma_{ 2 \, 2 }^{ 1  } =  -r + 2 \, M(u) \, , ~ \Gamma_{ 3 \, 3 }^{ 1 }  =  -r \sin^2 \theta + 2 \, M(u) \sin^2 \theta \, , \\
\Gamma_{ 1 \, 2 }^{ 2 }  =  \frac{1}{r} \, ,~ \Gamma_{ 3 \, 3 }^{ 2 } =  -\sin \theta \, \cos \theta \, ,~ \Gamma_{ 1 \, 3 }^{ 3  } =  \frac{1}{r} \, ,~ \Gamma_{ 2 \, 3 }^{ 3  } =  \frac{\cos \theta}{\sin \theta} \, .
\end{gather*}
%
The Weyl tensor has Petrov type D (see \cite{Petrov} for the Petrov classification of the Weyl tensor in terms of the multiplicities of its principal null directions), i.e. it has two double principal null directions that are given by
\begin{equation} \label{PNDs}
V = \frac{\partial}{\partial r} \, ,~ W = \frac{\partial}{\partial u} - \frac{1}{2} F \frac{\partial}{\partial r} \, .
\end{equation}
This is well known (see \cite{GriPo}) and can be checked easily by observing that $V$ and $W$ both satisfy the condition ensuring that they are at least double roots of the Weyl tensor (see R. Penrose, W. Rindler \cite{PeRi} Vol. 2, p. 224)
\[ C_{abc[d} V_{e]} V^b V^c = C_{abc[d} W_{e]} W^b W^c =0 \, .\]
We consider a null tetrad built using the principal null vectors above
\begin{eqnarray*}
l &=& V \, , \\
n &=& W \, , \\
m &=& \frac{1}{r\sqrt{2}} \left( \frac{\partial}{\partial \theta} + \frac{i}{\sin \theta} \frac{\partial}{\partial \varphi} \right) \, ,\\
\bar{m} &=& \frac{1}{r\sqrt{2}} \left( \frac{\partial}{\partial \theta} - \frac{i}{\sin \theta} \frac{\partial}{\partial \varphi} \right) \, .
\end{eqnarray*}
It is a normalised Newman-Penrose tetrad, i.e.
\[ l_a l^a = n_a n^a = m_a m^a = \bar{m}_a \bar{m}^a = l_a m^a = n_a m^a = 0 \, ,~ l_a n^a = - m_a \bar{m}^a = 1 \, .\]
Let $\{ o^A , \iota^A  \}$ be the spin-frame (a local basis of the spin-bundle $\S^A$ that is normalised, i.e. $o_A \iota^A =1$) defined uniquely up to an overall sign by
\[ l^a = o^A \bar{o}^{A'} \, ,~ n^a = \iota^A \bar{\iota}^{A'} \, ,~ m^a = o^A \bar{\iota}^{A'} \, ,~ \bar{m}^a = \iota^A \bar{o}^{A'} \, .\]
Since the spacetime has Petrov type D, the Weyl spinor $\Psi_{ABCD}$ has only one non-zero component which is
\[
\Psi_2 = \Psi_{ABCD} \, o^A o^B \iota^C \iota^D = -\frac{M(u)}{r^3} \, .
\]
%
%
%
%
%
%
%
%
The Ricci tensor is non-zero
\[ \newcommand{\Bold}[1]{\mathbf{#1}}\mathrm{Ric}\left(g\right) = -\frac{2 \, \dot{M}(u)}{r^{2}} \mathrm{d} u\otimes \mathrm{d} u \]
but trace-free, i.e.
\[ \mathrm{Scal}_g = 0 \, ,\]
and the only non-zero Newman-Penrose scalar for the Ricci tensor is
\[ \Phi_{22} = \frac12 R_{ab} n^a n^b = -\frac{\dot{M}(u)}{r^{2}} \, . \]
The curvature scalar, or Kretschmann scalar, is the total contraction of the Riemann tensor with itself. It is related to the analogous invariant for the Weyl tensor by the following formula (see C. Cherubini, D. Bini, S. Capozziello, R. Ruffini \cite{CheBiCaRu})
\[ k := R_{abcd} R^{abcd} = C_{abcd} C^{abcd} + 2 R_{ab} R^{ab} - \frac13 \mathrm{Scal}_g^2 \, .\]
For Vaidya's spacetime, we have
\begin{equation} \label{K}
k = R_{abcd} R^{abcd} =C_{abcd} C^{abcd}= \frac{48 M(u)^2}{r^6} \, .
\end{equation}
The metric \eqref{VaidyaMetric} is defined on $\R_u \times ]0,+\infty[_r \times S^2_\omega$ and describes a radiative white hole whose mass varies with time as a result of outgoing radiation carried by null dust. It is therefore natural to assume that $M (u)$ is a non-increasing function of $u$; this amounts to assuming that the null dust has positive energy. Another natural assumption is that the mass has finite limits as $u$ tends to $\pm \infty$:
\begin{equation} \label{Assumption1}
\lim_{u\rightarrow \pm \infty} M(u) \rightarrow M_\pm ~\mbox{with } 0 \leq M_+ < M_- < +\infty \, .
\end{equation}
In the case where $M$ is constant on $] -\infty , u_-] $ and on $[u_+ , +\infty [$ with $-\infty < u_- < u_+ <+\infty$, we have a Schwarzschild white hole of mass $M_-$ which emits a burst of null radiation between the retarded times $u_-$ and $u_+$ and eventually stabilises to a Schwarzschild black hole of mass $M_+ < M_-$ (unless $M_+=0$, in which case the white hole evaporates completely in finite time). If $M_+ >0$, the future event horizon is at $r = 2M_+$ but the location of the past horizon is not so clear. For $u<u_-$, it is located at $r=2M_-$ and it is a null hypersurface with spherical symmetry. Therefore, it is the hypersurface generated by the family of curves indexed by $\omega \in \s^2$:
\begin{equation} \label{gammau}
\gamma (u) = (u , r = r (u) , \omega ) \, ,~ u\in \R \, ,
\end{equation}
that are such that
\[ r(u) = 2M_- \mbox{ for } u \leq u_- \]
and have the property of being null, i.e.
\begin{equation} \label{ODE1NullGammadot}
g(\dot{\gamma} (u) , \dot{\gamma}(u) ) = 1- \frac{2M(u)}{r (u)} + 2\dot{r} (u) = 0 \, .
\end{equation}
Hence, the function $r(u)$ satisfies the following ordinary differential equation
\begin{equation} \label{ODE1}
\dot{r} (u) = -\frac{1}{2} \left( 1- \frac{2M(u)}{r (u)} \right) \, ,
\end{equation}
with $r>0$ and $r(u) = 2M_-$ for $u\leq u_-$.
\begin{remark}
We shall often (starting immediately below) identify the solutions to \eqref{ODE1} with the curves \eqref{gammau} satisfying \eqref{ODE1NullGammadot} and simply refer to the integral lines of \eqref{ODE1}.
\end{remark}
If we no longer assume that $M(u) = M_-$ in a neighbourhood of $-\infty$, the past horizon will be spanned by solutions to \eqref{ODE1} such that $r>0$ and $\lim_{u\rightarrow -\infty} r (u) = 2M_-$.
%
%

The ODE \eqref{ODE1} is in fact the general equation for a null curve that is transverse to the level hypersurfaces of $u$ (i.e. to $\nabla u$, that is a normal and tangent vector field to these hypersurfaces) and orthogonal to the orbits of the rotation Killing vectors. Vaidya's spacetime comes equipped with a null congruence, given by the lines of constant $u$ and $\omega$, which are the integral lines of
\begin{equation} \label{Gradu}
\nabla u = g^{-1} (\d u) = \frac{\partial}{\partial r} = V \, ,
\end{equation}
where $g^{-1}$ is the inverse Vaidya metric given by
\begin{equation} \label{InvMet}
g^{ab} \partial_a \partial_b = 2 \partial_u \partial_r - \left( 1 - \frac{2M(u)}{r} \right) \partial_r^2 -  r^{-2} \partial_\omega^2 \, ,
\end{equation}
$\partial_\omega^2$ denoting the euclidean inverse metric on $\s^2$. The integral lines of \eqref{ODE1} provide us with a second null congruence that is transverse to the first one, corresponding to the lines of constant $v$ and $\omega$ in the case of the Schwarzschild metric.
\begin{remark}
Note that the tangent vector to the integral curves of \eqref{ODE1} is exactly
\[ \frac{\partial}{\partial u} - \frac{1}{2} F \frac{\partial}{\partial r} =W \, .\]
The integral curves of \eqref{ODE1} are therefore the integral lines of the principal null vector field $W$. Since the spacetime is not vacuum, we do not have the Goldberg-Sachs Theorem that would ensure that these are geodesics. However we shall see in Subsection \ref{Optical} Proposition \ref{PNG2} that these curves are indeed geodesics; they are the family of incoming principal null geodesics and form the second natural null congruence of Vaidya's spacetime. Similarly the integral lines of $V$ are also geodesics (see Proposition \ref{PNG1}), they are the outgoing principal null geodesics of Vaidya's spacetime.
\end{remark}

\section{The incoming principal null congruence} \label{Incoming}

In this section, we analyse the qualitative behaviour of solutions to Equation \eqref{ODE1}, with special emphasis on the solutions generating the past horizon.
Our main results are proved under the assumption that
\begin{equation} \label{Assumption2}
\dot{M}(u) <0 \mbox{ on } ]u_-,u_+[\, ,~ -\infty \leq u_- < u_+ \leq +\infty \, ,~ \dot{M} \equiv 0\mbox{ elsewhere.}
\end{equation}
This covers the cases where the mass decreases strictly for all retarded times and where it decreases only on a finite retarded time interval. We dismiss as physically irrelevant the cases where intervals in $u$ with constant mass alternate with intervals on which the mass decreases. We also ignore, for similar reasons, cases where $\dot{M}$ vanishes at isolated points.

\subsection{General properties}

We start with an obvious observation.
\begin{lemma}
On an interval $]u_0,u_1[$ on which $\dot{M}(u)$ does not vanish everywhere, $r(u)$ cannot be identically equal to $2M(u)$.
\end{lemma}
{\bf  Proof.} If $r(u) = 2M(u)$ satisfies \eqref{ODE1} on $]u_0,u_1[$, then
\[ 2 \dot{M} (u) = -\frac12 \left( 1 - \frac{2 M(u)}{2M(u)}\right) = 0 \mbox{ on } ]u_0,u_1[ \, , \]
which contradicts the assumption. \qed

Then, we give an important estimate that is a consequence of the local uniqueness of solutions to the Cauchy problem for \eqref{ODE1}.
\begin{lemma} \label{Barrier}
Let $(]u_1 , u_2[ , r)$ be a solution to \eqref{ODE1} such that, for a given $u_0 \in ]u_1 , u_2[$, we have $r(u_0) \geq 2M(u_0)$. Let us assume that $\dot{M} (u) <0$ for all $u\in ]u_1 , u_2[$, then $r(u) > 2M(u)$ on $]u_0 , u_2[$.
\end{lemma}
{\bf Proof.} First, note that if $r(u_0) = 2M(u_0)$, then $\dot{r} (u_0) =0$, while $\dot{M}(u_0) <0$, hence there exists $\varepsilon >0$ such that in $]u_0 , u_0+\varepsilon[$ we have $r(u) > 2M(u)$. If $r(u_0) > 2M(u_0)$, then we have the same conclusion by continuity.

Now, let $u_3$ be the lowest value of $u$ in $]u_0,u_2[$ such that $r(u) = 2M(u)$. Then \eqref{ODE1} implies that $\dot{r} (u_3) = 0 > 2\dot{M} (u_3)$ and therefore, there exists $\delta >0$ such that $r(u) < 2M(u) $ in $]u_3-\delta , u_3[$. By continuity of $r$ and $M$, there exists $u_4 \in ] u_0 , u_3 [$ such that $r(u_4) = 2M (u_4)$. This contradicts the assumptions on $u_3$. It follows that $r(u) >2M(u)$ on $]u_0, u_2 [$. \qed

The asymptotic behaviour of maximal solutions to \eqref{ODE1} in the past is unstable. One solution has a finite limit $2M_-$; it corresponds to the past event horizon. All other solutions either end at past null infinity or reach the past singularity in finite retarded time. The following theorem gives a complete classification of the solutions to \eqref{ODE1} in terms of their behaviour in the past and also describes precisely their behaviour in the future.
\begin{theorem} \label{ThmHorizon}
Under Assumptions \eqref{MSmooth}, \eqref{Assumption1} and \eqref{Assumption2}, there exists a unique maximal solution $r_h$ to \eqref{ODE1} such that
    \[ \lim_{u\rightarrow -\infty} r_h(u) = 2M_- \, .\]
\begin{itemize}
\item If either $M_+ >0$ or $u_+=+\infty$, $r_h$ exists on the whole real line, $r_h (u) \rightarrow 2M_+$ as $u\rightarrow +\infty$ and any other maximal solution $r$ to \eqref{ODE1} belongs to either of the following two categories:
    \begin{enumerate}
    \item $r$ exists on the whole real line, $r(u) > r_h (u)$ for all $u \in \R$, $\lim_{u\rightarrow -\infty} r(u) = +\infty$ and
    $\lim_{u\rightarrow +\infty} r(u) = 2M_+$;
    \item $r$ exists on $]u_0 , +\infty [$ with $u_0 \in \R$ and satisfies:  $r(u) < r_h (u) $ for all $u \in ]u_0 , +\infty[$, $\lim_{u\rightarrow u_0} r(u) = 0$ and $\lim_{u\rightarrow +\infty} r(u) = 2M_+$.
    \end{enumerate}
\item If $M_+=0$ and $u_+ <+\infty$, $r_h$ exists on an interval $]-\infty , u_0[$ with $u_+ \leq u_0 <+\infty$ and $\lim_{u\rightarrow u_0} r_h (u) = 0$. The other maximal solutions are of two types:
\begin{enumerate}
\item $r$ exists on $]-\infty , u_1[$ with $u_0 \leq u_1 < +\infty$, $r(u) > r_h (u)$ on $]-\infty , u_0[$, $\lim_{u \rightarrow u_1}r(u) = 0$ and $\lim_{u\rightarrow -\infty} r(u) = +\infty$;
\item $r$ exists on $]u_1 , u_2[$ with $-\infty < u_1 < u_2 \leq u_0$, $r(u) \rightarrow 0$ as $u$ tends to either $u_1$ or $u_2$ and $r(u) < r_h (u)$ on $]u_1 , u_2[$.
\end{enumerate}
\end{itemize}
\end{theorem}
{\bf Proof.} \begin{description} \item[Step 1: uniqueness of a maximal solution with finite limit as $u\rightarrow -\infty$.]
First, if a solution $r$ exists on an interval of the form $]-\infty , u_0[$ and has a finite limit at $-\infty$, then this limit must be $2M_-$. Indeed let us denote this limit by $l$, using \eqref{ODE1},
\[ \lim_{u \rightarrow -\infty } \dot{r} (u) = -\frac{1}{2} \left( 1 - \frac{2M_-}{l} \right) \, . \]
So $\dot{r}$ also has a finite limit at $-\infty$ and this limit must be zero in order not to contradict the finite limit of $r(u)$, i.e. $l=2M_-$.

Then let us show that there is at most one solution to \eqref{ODE1} defined on an interval of the form $]-\infty , u_0 [$ such that
\[ \lim_{u\rightarrow -\infty} r(u) = 2M_- \, . \]
Let us assume that there are two such solutions $r_1$ and $r_2$. Then $\psi = r_2-r_1$ satisfies
\begin{equation} \label{EqDiff}
\dot{\psi} (u) = \frac{M}{r_2} - \frac{M}{r_1} = \frac{-M}{r_1r_2} \psi 
\end{equation}
and
\[ \lim_{u\rightarrow -\infty} \psi (u) =0 \, . \]
However, since
\[ \lim_{u \rightarrow -\infty} \frac{-M}{r_1r_2} = \frac{-1}{4M_-} <0 \, ,\]
if follows that unless $\psi$ is identically zero,
\[ \left( \log (\vert \psi \vert ) \right)' \longrightarrow \frac{-1}{4M_-} \mbox{ as } u \rightarrow -\infty \]
and $\psi$ blows up exponentially fast at $-\infty$. Since we know that $\psi$ tends to zero at $-\infty$, we conclude that $\psi$ is identically zero, i.e. $r_1=r_2$.
\item[Step2: construction of the past horizon.] Now we construct a solution to \eqref{ODE1} that tends to $2M_-$ at $-\infty$.
Let us first consider the case where $u_- = -\infty$. For each $n \in \N$, we define $r_n$ to be the maximal solution to \eqref{ODE1} such that $r_n (-n) = 2M(-n)$. It exists on an interval of the form $]u^1_n, u^2_n [$, $u^1_n <-n < u^2_n$. Let $u^3_n = \min \{ u^2_n , u_+\}$. By Lemma \ref{Barrier}, $r_n (u) >2M(u)$ on $]-n,u^3_n[$, hence $\dot{r}_n <0$ there and it follows that $r_n (u) < 2M (-n)$. These a priori bounds imply that $u_n^2 \geq u_+$. Therefore, $u^2_n = +\infty$ in the case where $u_+ = +\infty$. If $u_+<+\infty$ and $M_+>0$, then we could have $r_n(u_+)=2M_+$, in which case $u_n^2=+\infty$ and $r_n (u) = 2M_+$ for $u\geq u_+$, or $r_n(u_+) > 2M_+$ and then $r_n (u) > 2M_+$ on $[u_+ , u_n^2[$ since two solutions cannot cross, whence $\dot{r}_n$ is negative there and we infer $u_n^2=+\infty$. If $u_+<+\infty$ and $M_+=0$ then on $]u_+, u_n^2[$, $r_n$ satisfies the simple ODE
\[ \dot{r}_n = -\frac12 \, ,\]
and $r_n$ reaches $0$ in finite retarded time. Hence in this case $u_n^2 <+\infty$.

Using again the fact that, by uniqueness of solutions to the Cauchy problem for \eqref{ODE1}, two solutions cannot cross, we infer that the sequence $u_n^2$ is increasing. Let
\[ u^2 = \lim_{n\rightarrow +\infty} u^2_n \, .\]
For any compact interval $I$ of $]-\infty , u^2[$, there exists $n_0 \in \N$ such that the sequence $(r_n )_{n \geq n_0}$ is well-defined, increasing and bounded on $I$. Hence, Lebesgue's dominated convergence theorem implies that the sequence $(r_n)$ converges in $L^1_{\mathrm{loc}} (]-\infty , u^2[)$ towards a positive function $r_h$ such that
\begin{equation} \label{Sandwichrh}
2M(u) \leq r_h (u) \leq 2 M_- ~\forall u \in ]-\infty , u^2[ \, .
\end{equation}
Moreover, $1/r_n$ also converges towards $1/r_h$ in $L^1_{\mathrm{loc}} (]-\infty , u^2[)$, because, for any given compact interval $I$, it is a well-defined, decreasing and bounded sequence on $I$ for $n$ large enough. This implies, by equation \eqref{ODE1} for $r_n$, that $\dot{r}_n$ converges  in $L^1_\mathrm{loc} (]-\infty , u^2[)$ and by uniqueness of the limit in the sense of distributions, the limit must be $\dot{r}_h$. Consequently $r_h$ is a solution to \eqref{ODE1} in the sense of distributions. An easy bootstrap argument then shows that $r_h$ is a strong solution to \eqref{ODE1} and is in fact smooth on $]-\infty , u^2[$.
    
Besides, by \eqref{Sandwichrh} and the fact that $M(u ) \rightarrow M_-$ as $u \rightarrow -\infty$, it follows that
\[ \lim_{u\rightarrow -\infty} r_h (u) = 2M_- \, . \]
In the case where $u_- >-\infty$, we simply need to consider the maximal solution to \eqref{ODE1} such that $r(u_- -1) = 2M_-$. This solution exists on an interval of the form $]-\infty , u^2[$ and satisfies \eqref{Sandwichrh}.

Let us now turn to the value of $u^2$ and the behaviour of $r_h$ in the future. If either $M_+>0$ or $u_+=+\infty$, then by \eqref{Sandwichrh}, we have $u^2 =+\infty$. By \eqref{Sandwichrh} again, $\dot{r}_h(u) <0$ on $\R$ and it follows that $r_h (u) $ has a finite limit $l$ as $u \rightarrow +\infty$. If $M_+>0$, we have
\[ \dot{r}_h (u) \rightarrow -\frac12 \left( 1 - \frac{2M_+}{l} \right) \mbox{ as } u\rightarrow +\infty \]
and we must have $l=2M_+$ or contradict the finite limit of $r_h$. If $M_+=0$ and $u_+=+\infty$, then if $l\neq 0$,
\[ \dot{r}_h (u) \rightarrow -\frac12 \mbox{ as } u\rightarrow +\infty \]
which is incompatible with $u^2=+\infty$. Finally, if $M_+=0$ and $u_+<+\infty$, then \eqref{Sandwichrh} implies that $u^2 \geq u^+$. If $r_h(u_+)=0$ then the solution terminates at $u=u_+$, $u^2 = u_+$. Otherwise, $u^2 >u_+$ and on $[u_+ , u^2[$ we have
\[ \dot{r}_{h} (u)= -\frac12 \, ,\]
as long as $r_h(u)$ remains positive. Therefore, we have
\[ r_h (u) = r_h (u_+) - (u-u_+) \mbox{ for } u_+ \leq u \leq u_++r_h (u_+ ) \]
and the integral curve ends at $u=u_++r_h (u_+ )$, i.e. $u^2$ is finite and is equal to $u_++r_h (u_+ )$. Hence for $M_+=0$ and $u_+<+\infty$, the past event horizon vanishes in finite retarded time $u_++r_h(u_+)$ and there is no future event horizon.
\item[Step 3: classification of the other maximal solutions.]
\mbox{}
\begin{itemize}
\item We begin with the case where either $u_+=+\infty$ or $M_+>0$.
Let $(]u_1,u_2[,r)$ be a maximal solution to \eqref{ODE1}. Let $u_0 \in ]u_1,u_2[$ and assume that $0< r(u_0) < r_h (u_0)$ (resp. $r(u_0)>r_h(u_0)$). By uniqueness of solutions to the Cauchy problem for \eqref{ODE1}, solutions cannot cross, so for all $u \in ]u_1,u_2[$ we have $0< r(u) < r_h (u)$ (resp. $r(u)>r_h(u)$).
\begin{enumerate}
\item Case where $0< r(u_0) < r_h (u_0)$. Let us first assume that $r(u_0) > 2M(u_0)$. If $r(u) > 2M(u)$ on its interval of existence, then $r(u)$ is bounded between $2M(u)$ and $r_h(u)$ and we must have $]u_1,u_2[\, =\R$. However, we then have $r(u) \rightarrow 2M_-$ as $u\rightarrow -\infty$ and this contradicts the uniqueness of $r_h$. It follows that there exists $u_3 \in ]u_1,u_2[$ such that $r(u_3) = 2M(u_3)$. Therefore $r(u) < 2M(u)$ on $]u_1,u_3[$ (the proof is similar to that of Lemma \ref{Barrier}), $r$ is an increasing function on this interval and $\dot{r}$ is decreasing.  This implies that $r(u)$ must reach $0$ in finite time in the past and keep on existing towards the past as long as it has not reached $0$. Hence $u_1 >-\infty$ and $r(u) \rightarrow 0$ as $u\rightarrow u_1$. Since $r(u_3) = 2M (u_3)$, then we have by Lemma \ref{Barrier} that $r(u) > 2M(u)$ on $]u_3,u_+[$ and since solutions do not cross, $r(u) \geq 2M(u)$ on $[u_+,u_2[$. Hence
\[ 2M(u) \leq r(u) < r_h (u) \mbox{ on } ]u_0,u_2[ \, .\]
This implies that $u_2=+\infty$ and $\lim_{u\rightarrow +\infty} r(u) = 2M_+$.

If $r(u_0) = 2M (u_0)$, then we can repeat the arguments above, replacing $u_3$ by $u_0$; we infer: $u_1 >-\infty$ and $\lim_{u\rightarrow u_1} r(u) = 0$, $u_2=+\infty$ and $\lim_{u\rightarrow +\infty} r(u) = 2M_+$.

If $r(u_0) < 2M(u_0)$ then $r(u)$ increases as long as $r(u) < 2M(u)$. Either $r(u) < 2M(u)$ on its whole interval of existence (note that this requires $M_+>0$), in which case $u_2=+\infty$, or there exists $u_4\in ]u_0,u_2[$ such that $r(u_4)=2M(u_4)$ and we can then use the same reasoning as before on $]u_4,u_2[$ and infer that $u_2=+\infty$. In the latter case, we have as before $\lim_{u\rightarrow +\infty} r(u) = 2M_+$. In the former, $r(u)$ has a finite positive limit $l$ as $u\rightarrow +\infty$ and (recall that we must have $M_+>0$)
\[ \dot{r}(u) \rightarrow -\frac12 \left( 1 - \frac{2M_+}{l} \right) \mbox{ as } u\rightarrow +\infty \]
and we must have $l=2M_+$ in order not to contradict $u_2=+\infty$. In both cases, we have $u_1 >-\infty$ and $r(u) \rightarrow 0$ as $u\rightarrow u_1$.
\item If $r(u_0)>r_h(u_0)$ then $r$ is a decreasing function on its interval of existence and is bounded below by $r_h$. This implies that $u_2=+\infty$. Moreover, on its whole interval of existence, $r$ satisfies
\[ -\frac12 < \dot{r} (u) <0 \]
and it follows that $u_1=-\infty$. Since $r$ is a decreasing function on $\R$, it has a limit as $u \rightarrow -\infty$ and we have seen above that this limit cannot be finite, hence
\[ \lim_{u\rightarrow -\infty} r(u) = +\infty \, .\]
The solution $r$ also has a limit $l$ as $u\rightarrow +\infty$ and since $r$ is a decreasing function and $r(u) > r_h (u) >2M(u)$ on $\R$, it follows that $2M_+ \leq l <+\infty$. If $M_+ >0$, then $l>0$ and $\dot{r}$ also has a limit as $u\rightarrow +\infty$ given by
\[ \lim_{u\rightarrow+\infty} \dot{r}(u) = -\frac12 \left( 1 - \frac{2M_+}{l} \right)\, . \]
This must be zero in order not to contradict the finite limit of $r$. Hence $l =2M_+$. If $M+=0$, then unless $l=0$ we have that
\[ \lim_{u\rightarrow+\infty} \dot{r}(u) = -\frac12\]
which implies that $r(u)$ must reach $0$ in finite retarded time and contradicts $u_2=+\infty$. Hence in this case we have $l=0=2M_+$.
\end{enumerate}
\item In the case where $M_+=0$ and $u_+<+\infty$, the proof uses exactly the same arguments as in step 3 and the end of step 2. \qed
\end{itemize}
\end{description}

If the mass decreases only for a finite range of $u$, we have not been able to rule out cases for which we have $r_h(u_+)=2M_+$, nor have we managed to find explicit examples of this situation. It is however easy to see that there are cases for which $r_h(u) > 2M_+$ for all $u \in \R$.
\begin{proposition}
In the case where $u_\pm$ are both finite, assume that $u_+ - u_- < 4 (M_- - M_+)$, then $r_h(u_+) > 2M_+$ and therefore $r_h(u) > 2M_+$ for all $u \in \R$.
\end{proposition}
{\bf Proof.} It is a simple observation. We have
\begin{eqnarray*}
r_h(u_+) - r_h(u_-) &=& \int_{u_-}^{u_+} \dot{r}_h(u) \d u \\
&=& -\frac12 \int_{u_-}^{u_+} \left( 1 - \frac{2M(u)}{r_h(u)} \right) \d u \\
&=& -\frac12 (u_+-u_-) + \int_{u_-}^{u_+} \frac{M(u)}{r_h(u)} \d u > -\frac12 (u_+-u_-) \, .
\end{eqnarray*}
Since $r_h(u_-) = 2M_-$,
\[ r(u_+) > 2M_- -\frac12 (u_+-u_-) >2M_+\, . \]
Since on $]u_+,+\infty[$, $2M_+$ is a solution to \eqref{ODE1}, then by uniqueness of solutions we must have $r_h (u) >2M_+$ for all $u>u_+$. This proves the proposition. \qed

\subsection{The second optical function} \label{Optical}

The function $u$ is an optical function, which means that its gradient is a null vector field, or equivalently that $u$ satisfies the eikonal equation
\begin{equation} \label{Eikonalu}
g (\nabla u , \nabla u ) =0 \, .
\end{equation}
An important property of optical functions is that the integral lines of their gradient are null geodesics with affine parametrisation. This is established in \cite{HaNi}. The more complete Propositions (7.1.60) and (7.1.61) in Penrose and Rindler Vol 2 \cite{PeRi} state that for a null congruence, the following three properties are equivalent~:
\begin{enumerate}
\item it is hypersurface-orthogonal;
\item it is hypersurface-forming;
\item it is geodetic and twist-free.
\end{enumerate}
We recall the proof of the fact that the integral curves of an optical function are null geodesics, as it is a straightforward calculation.
\begin{lemma} \label{NablaxiGeod}
Let $\xi$ be an optical function and denote $\cal L = \nabla \xi$. The integral curves of $\cal L$ are geodesics and $\cal L$ corresponds to a choice of affine parameter, i.e.
\[\nabla_{\cal L} {\cal L} = 0 \, .\]
\end{lemma}
{\bf Proof.} The proof is direct~:
\begin{eqnarray*}
\nabla_{\cal L} {\cal L}^b &=& \nabla_{\nabla \xi} {\nabla^b \xi} \, , \\
&=& \nabla_a \xi \nabla^a \nabla^b \xi \, ,\\
&=& \nabla_a \xi \nabla^b \nabla^a \xi \mbox{ since the connection is torsion-free,} \\
&=& \nabla^b \left( \nabla_a \xi \nabla^a \xi \right) - \left( \nabla^b \nabla_a \xi \right) \nabla^a \xi \, ,\\
&=& 0 - \nabla_a \xi \nabla^a \nabla^b \xi \mbox{ since } \nabla \xi \mbox{ is null and the connection torsion-free,} \\
&=& - \nabla_{\nabla \xi} \nabla^b \xi  \, . \qed
\end{eqnarray*}
Since (see \eqref{Gradu})
\[ \nabla u = \frac{\partial}{\partial r} = V \]
and $V$ is a principal null direction of the Weyl tensor, a consequence of Lemma \ref{NablaxiGeod} and of \eqref{Eikonalu}  is the following.
\begin{proposition} \label{PNG1}
The integral lines of $V$ are affinely parametrised null geodesics; they are the outgoing principal null geodesics of Vaidya's spacetime.
\end{proposition}
We now establish the existence of a second optical function.
\begin{proposition} \label{PNG2}
There exists a function $v$ defined on $\R_u \times ]0,+\infty[_r \times S^2_{\omega}$, depending solely on $u$ and $r$, such that $\nabla v$ is everywhere tangent to the integral lines of \eqref{ODE1}. This means that $g(\nabla v , \nabla v )=0$, i.e. $v$ is an optical function. The integral lines of \eqref{ODE1}, which are also the integral lines of $\nabla v$, are therefore null geodesics and their congruence generates the level hypersurfaces of $v$. Since the integral lines of \eqref{ODE1} are also tangent to $W$ (defined in \eqref{PNDs}) it follows that they are the incoming principal null geodesics of Vaidya's spacetime.
\end{proposition}
{\bf Proof.} The metric $g$ can be written as
\[ g = F \d u \left( \d u + 2 F^{-1} \d r \right) - r^2 \d \omega^2 \, .\]
Following the construction of $v$ for the Schwarzschild metric, it is tempting to put
\[ \d v = \d u + 2 F^{-1} \d r = 2F^{-1} g^{-1} (W)\, ,\]
however this $1$-form is not closed since
\[ \d \left( \d u + 2 F^{-1} \d r \right) = -2 F^{-2} \frac{\partial F}{\partial u} \d u \wedge \d r = \frac{4\dot{M}(u)}{rF^2} \d u \wedge \d r \]
which vanishes identically only if the mass $M$ is constant, i.e. in the Schwarzschild case. We introduce an auxiliary function $\psi >0$ and we write
\[ g = \frac{F}{\psi} \d u \left( \psi \d u + 2 \psi F^{-1} \d r \right) - r^2 \d \omega^2 \, .\]
Our purpose is to find conditions on $\psi$ that ensure that the $1$-form $\alpha := \psi \d u + 2 \psi F^{-1} \d r$ is exact. Since we work in the variables $(u,r)$ on the simply connected domain $\R_u \times ]0,+\infty[_r$, all that is required is that $\alpha$ be closed, i.e. that
\[ \d \alpha = 2 \frac{\partial}{\partial u} \left( \frac{\psi}{F} \right) - \frac{\partial \psi}{\partial r} = 0 \, .\]
This equation has the more explicit form
\begin{equation} \label{PDE1}
\frac{\partial \psi}{\partial u} - \frac{F}{2} \frac{\partial \psi}{\partial r} + \frac{2}{F} \frac{\dot{M}}{r} \psi = 0 \, .
\end{equation}
This is an ordinary differential equation along the integral lines of the second principal null direction (defined in \eqref{PNDs})
\[ W = \frac{\partial}{\partial u} - \frac{F}{2} \frac{\partial}{\partial r} \]
parametrised by $u$. Let $\gamma (u) = (u,r(u),\omega)$ be an integral line of $W$ (which is equivalent to $r(u)$ being a solution to \eqref{ODE1}), Equation \eqref{PDE1} along $\gamma$ reads
\[ \frac{\d}{\d u} (\psi \circ \gamma)= \left( -\frac{2\dot{M}}{rF} \psi \right) \circ \gamma \, ,\]
or equivalently
\begin{equation} \label{ODEpsi}
\frac{\d}{\d u} \left( \log \left\vert \psi \circ \gamma \right\vert \right)= \left( -\frac{2\dot{M}}{rF} \right) \circ \gamma \, .
\end{equation}
Equation \eqref{PDE1} can therefore be integrated as follows. First, we take a hypersurface transverse to all the integral lines of \eqref{ODE1}, for instance $\mathcal{S} = \{ u=0 \}$ and, we fix the value of $\psi$ on $\mathcal{S}$, say $\psi =1$ on $\mathcal{S}$. Then we evaluate $\psi$ on each integral line of \eqref{ODE1} by solving the ODE \eqref{ODEpsi}. Since the integral lines of \eqref{ODE1} are a congruence of $\R_u \times ]0,+\infty[_r\times S^2_\omega$, this allows to define $\psi$ on this whole domain as a smooth (by smooth dependence on initial data) and nowhere vanishing function. The $1$-form $\alpha$ is then closed on $\R_u \times ]0,+\infty[_r\times S^2_\omega$. Since $\alpha$ depends only on $u$ and $r$, we may see it as a closed $1$-form on $\R_u \times ]0,+\infty[_r$ which is simply connected. It follows that $\alpha$ is exact on $\R_u \times ]0,+\infty[_r\times S^2_\omega$ and modulo a choice of hypersurface $\Sigma$ generated by the integral lines of \eqref{ODE1}, we can define a function $v$ on $\R_u \times ]0,+\infty[_r\times S^2_\omega$ such that $v=0$ on $\Sigma$ and $\alpha = \d v$. In particular,
\[ \d v = 2 \psi F^{-1} g^{-1} (W) \mbox{ and } \nabla v = 2\psi F^{-1} W \, . \qed\]

\section{Case of a complete evaporation in infinite time} \label{Evaporation}

We now devote particular attention to the case where $M_+ =0$ and $u_+=+\infty$. As before, we assume that $\dot{M}<0$ on $]u_-,+\infty[$.

\subsection{The asymptotic null singularity}

As we have established in Theorem \ref{ThmHorizon}, the past event horizon ends up at $r=0$ as $u \rightarrow +\infty$ and so do all the integral curves of \eqref{ODE1}, i.e. all the incoming principal null geodesics. From this, we infer the following theorem.
\begin{theorem} \label{NullSingularityK}
Whatever the speed at which $M(u) \rightarrow 0$ as $u \rightarrow +\infty$, we have a null singularity of the conformal structure in the future of our spacetime. More precisely, the Kretschmann scalar $k$ does not remain bounded as $u \rightarrow +\infty$ along any integral line of \eqref{ODE1}.
\end{theorem}
{\bf Proof.} Consider $(]u_0,+\infty [,r)$ a maximal solution to \eqref{ODE1}, with $u_0 \in \R \cup \{ -\infty \}$. Assume that $k$ remains bounded along the integral line as $u \rightarrow +\infty$. Then, using \eqref{K}, so does $M/r^3$ and it follows that $M/r$ tends to $0$ as $u\rightarrow +\infty$ along the integral line. This implies in turn that $\dot{r}(u) \rightarrow -1/2$ as $u\rightarrow +\infty$, which contradicts the fact that $r(u) \rightarrow 0$ as $u\rightarrow +\infty$. \qed

\begin{remark} \label{PreciseEquivrk}
If we assume that along the integral lines of \eqref{ODE1}, $\dot{r}(u)$ has a limit as $u\rightarrow +\infty$, this limit is necessarily zero in order not to contradict the fact that $r(u) \rightarrow 0$ as $u\rightarrow +\infty$. This implies in turn that along the integral line,
\[ \frac{M(u)}{r(u)} \rightarrow \frac12 \mbox{ as } u\rightarrow +\infty \, ,\]
i.e.
\begin{equation} \label{equivr}
r(u) \simeq 2M(u)  \mbox{ as } u\rightarrow +\infty
\end{equation}
and
\begin{equation} \label{Equivk}
k \simeq \frac{3}{4M(u)^4} \mbox{ as } u\rightarrow +\infty \, .
\end{equation}
%
%
\end{remark}

\subsection{A family of uniformly timelike congruences}

Some uniformly timelike curves also end up at $r=0$ as $u \rightarrow +\infty$. Let us consider a curve
\[ \gamma (u) = (u, r(u), \theta , \varphi ) \, ,\]
such that
\begin{equation} \label{EpsilonTimelike}
g (\dot{\gamma} (u) , \dot{\gamma} (u) ) = \varepsilon^2 \, ,~ \varepsilon >0 \, ,
\end{equation}
then $r$ satisfies the differential equation
\begin{equation} \label{ODE2}
\dot{r} (u) = \frac{\varepsilon^2}{2} - \frac12 \left( 1 - \frac{2M(u)}{r(u)} \right) \, .
\end{equation}
The tangent vector is
\[ \tau = \partial_u + \left( \frac{\varepsilon^2}{2} - \frac12 \left( 1 - \frac{2M(u)}{r} \right) \right) \partial_r \]
and
\[ \nabla_\tau \tau = -\frac{M(u)}{r^2} \left( \tau - \varepsilon^2  \partial_r \right) \, ,\]
so the integral curves of \eqref{ODE2} are not geodesics, except for $\varepsilon =0$. 
The behaviour of the integral curves of \eqref{ODE2} changes radically according to the value of $\varepsilon$. This is detailed in the next two propositions. The first one deals with the case where $0 < \varepsilon <1$.
\begin{proposition} \label{Epsilon<1}
Let $0<\varepsilon <1$ be given. There exists a unique maximal solution $r_\varepsilon$ to \eqref{ODE2} such that
\[ \lim_{u\rightarrow -\infty} r_\varepsilon (u) = \frac{2M_-}{1-\varepsilon^2} \, .\]
This solution exists on the whole real line and $r_\varepsilon (u) \rightarrow 0$ as $u\rightarrow +\infty$. Any other maximal solution $r$ to \eqref{ODE1} belongs to either of the two categories~:
\begin{enumerate}
\item $r$ exists on the whole real line, $r(u) > r_\varepsilon (u)$ for all $u \in \R$, $\lim_{u\rightarrow -\infty} r(u) = +\infty$ and $\lim_{u\rightarrow +\infty} r(u) = 0$~;
\item $r$ exists on $]u_0 , +\infty [$ with $u_0\in \R$ and satisfies~: $r(u) < r_\varepsilon (u)$ for all $u \in ]u_0 , +\infty[$ and $r(u)$ tends to $0$ as $u\rightarrow u_0$ and as $u\rightarrow +\infty$.
\end{enumerate}
Moreover, the Kretschmann scalar $k$ fails to be bounded on the integral lines of \eqref{ODE2} as $u\rightarrow +\infty$. If we assume moreover that $\dot{r}$ has a limit as $u\rightarrow +\infty$ along the integral lines of \eqref{ODE2}, then we have a similar behaviour for $k$ to that described in Remark \ref{PreciseEquivrk} for the integral lines of \eqref{ODE1}, namely
\begin{equation} \label{KTimelike}
k \simeq \frac{3(1-\varepsilon^2)^6}{4M(u)^4} \mbox{ as } u\rightarrow +\infty \, .
\end{equation}
\end{proposition}
The second proposition treats the cases where $\varepsilon \geq1$.
\begin{proposition} \label{EpsilonGeq1}
If $\varepsilon \geq 1$, then all maximal solutions to \eqref{ODE2} exist on a interval $]u_0 , +\infty[$ with $u_0 >-\infty$, $r$ is strictly increasing on $]u_0,+\infty[$ and $r(u) \rightarrow 0$ as $u\rightarrow u_0$. Moreover:
\begin{itemize}
\item if $\varepsilon =1$, then the limit of $r(u)$ as $u \rightarrow +\infty$ is finite if and only if $M(u)$ is integrable in the neighbourhood of $+\infty$;
\item if $\varepsilon >1$, then $r(u) \rightarrow +\infty$ as $u \rightarrow +\infty$.
\end{itemize}
\end{proposition}
\begin{remark}
In view of \eqref{EpsilonTimelike}, the proper time along an integral curve of \eqref{ODE2} is exactly given (after an adequate choice of origin) by $\tau = \varepsilon u$. Therefore, the behaviour of the integral lines of \eqref{ODE2} as $u \rightarrow +\infty$ described in Propositions \ref{Epsilon<1} and \ref{EpsilonGeq1} corresponds to their behaviour as proper time tends to $+\infty$.
\end{remark}
{\bf Proof of Proposition \ref{Epsilon<1}.} We observe that Equation \eqref{ODE2} can be transformed to \eqref{ODE1}. We put
\[ \tilde{r} (u) = \frac{r(u)}{1-\varepsilon^2} \, ,~ \tilde{M}(u) = \frac{M(u)}{(1-\varepsilon^2)^2} \, ,\]
then Equation \eqref{ODE2} becomes
\[ \dot{\tilde{r}}(u) = -\frac12 \left( 1 - \frac{2\tilde{M}(u)}{\tilde{r}(u)} \right) \, .\]
The classification of maximal solutions therefore follows directly from Theorem \ref{ThmHorizon}. The derivation of the behaviour of the Kretschmann scalar along an integral line is also similar to the null case. The proof of the lack of boundedness is the same as that of Theorem \ref{NullSingularityK} and assuming that $\dot{r}$ has a limit as $u\rightarrow +\infty$ along an integral line of \eqref{ODE2}, this limit must be zero and we infer
\[ r(u) \simeq \frac{2M(u)}{1-\varepsilon^2} \, .\]
Then \eqref{KTimelike} follows from \eqref{K}. \qed

{\bf Proof of Proposition \ref{EpsilonGeq1}.} Let $(]a,b[,r)$ be a maximal solution to \eqref{ODE2}.
\begin{itemize}
\item {\it Case where $\varepsilon =1$.} The differential equation \eqref{ODE2} becomes
\[ \dot{r} (u) = \frac{M(u)}{r(u)} \, ,\]
or equivalently
\[ \frac{\d}{\d u} ((r(u))^2) = 2M (u) \, .\]
The function $r$ is strictly increasing and given $u_1 \in ]a,b[$, we have for all $u \in ]a,b[$
\[ r(u)^2 = r(u_1)^2 + 2\int_{u_1}^u M(s) \d s \, .\]
Then $a$ is finite, strictly lower than $u_1$ and is precisely such that
\[ \int^{u_1}_a M(s) \d s = \frac{r(u_1)^2}{2} \, .\]
Also $b=+\infty$ and
\[ \lim_{u\rightarrow +\infty} r(u)^2 = r(u_1)^2 + 2\int_{u_1}^{+\infty} M(s) \d s \, .\]
\item {\it Case where $\varepsilon >1$.} Now for all $u\in ]a,b[$,
\begin{equation} \label{rdotpositive}
\dot{r}(u) > \frac{\varepsilon^2-1}{2} >0 \, .
\end{equation}
It follows that $a$ is finite and
\[ \lim_{u \rightarrow a} r(u) =0 \, .\]
Moreover, let $u_1 \in ]a,b[$, then using the fact that $r$ is strictly increasing on $]a,b[$, we have for all $u \in ]u_1,b[$,
\[ \dot{r}(u) <\frac{\varepsilon^2-1}{2} + \frac{M_+}{r(u_1)} \, ,\]
whence $b=+\infty$ and \eqref{rdotpositive} implies
\[ \lim_{u \rightarrow +\infty} r(u) = +\infty \, . \qed\]
\end{itemize}

\subsection{The global structure of the spacetime}

The two congruences of null geodesics that we have considered (the curves of constant $(u,\omega)$ and of constant $(v,\omega)$) are inextendible. The spacetime is therefore maximally extended and we have two global charts $\R_u \times ]0,+\infty[_r \times S^2_\omega$ and $\R_v \times ]0,+\infty[_r \times S^2_\omega$. Figures \ref{IncomingNull} to \ref{EpsGeq1} display the Penrose diagram of Vaidya's spacetime for a mass function that decreases strictly on the whole real line, tends to $0$ as $u\rightarrow +\infty$ and to a finite positive limit $M_+$ as $u \rightarrow -\infty$, with the general forms of various congruences: the incoming principal null geodesics (lines of constant $(v,\omega)$) in Figure \ref{IncomingNull}, the outgoing principal null geodesics (lines of constant $(u,\omega)$) in Figure \ref{OutgoingNull}, the timelike curves given by the integral lines of \eqref{ODE2} for $0<\varepsilon <1$ in Figure \ref{EpsInf1} and for $\varepsilon \geq 1$ in Figure \ref{EpsGeq1}. The dashed lines are curvature singularities. The null singularity in the future is an asymptotic singularity.
\begin{center}
\begin{figure}[ht!]
\begin{minipage}{7.5cm}
\includegraphics[width=7.5cm]{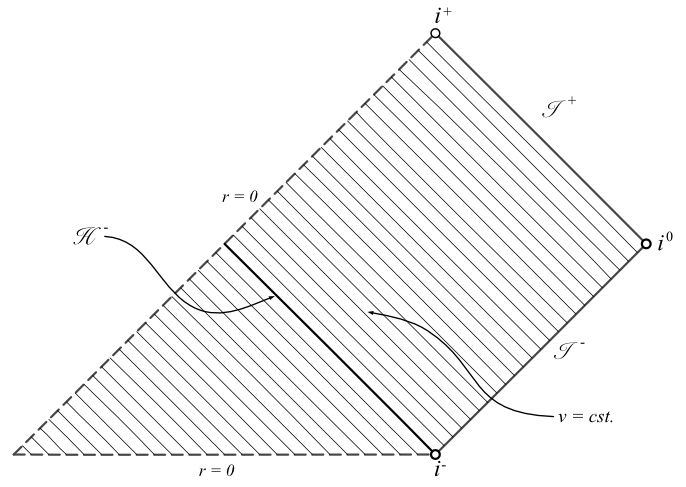}
\caption{Incoming principal null congruence: lines of constant $(v,\omega)$.}
\label{IncomingNull}
\end{minipage}
\hspace*{0.5cm}
\begin{minipage}{7.5cm}
\includegraphics[width=7.5cm]{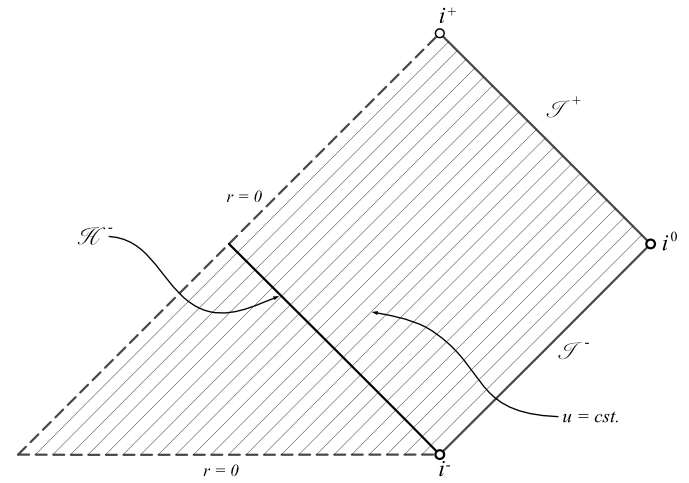}
\caption{Outgoing principal null congruence: lines of constant $(u,\omega)$.}
\label{OutgoingNull}
\end{minipage}
\end{figure}
\begin{figure}[ht!]
\begin{minipage}{7.5cm}
\includegraphics[width=7.5cm]{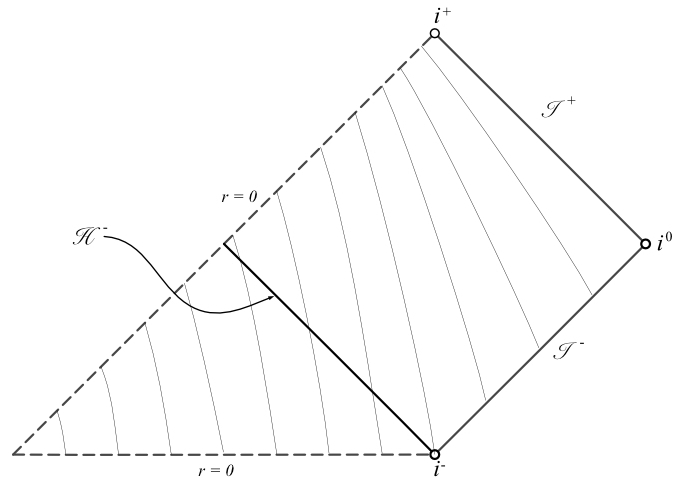}
\caption{The timelike congruence for $\varepsilon <1$.}
\label{EpsInf1}
\end{minipage}
\hspace*{0.5cm}
\begin{minipage}{7.5cm}
\includegraphics[width=7.5cm]{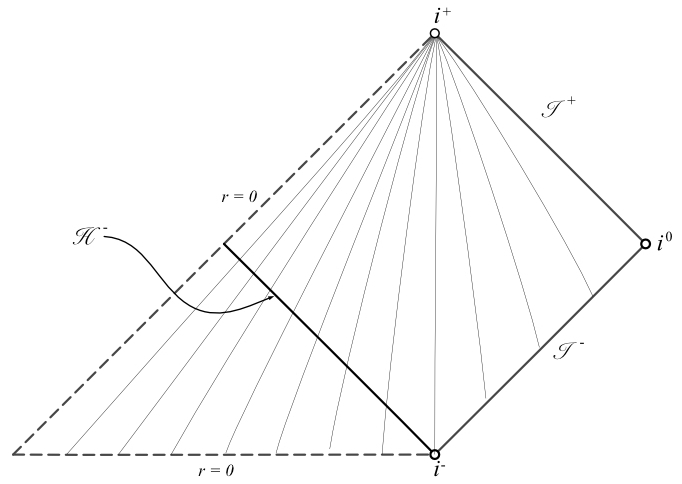}
\caption{The timelike congruence for $\varepsilon \geq 1$.}
\label{EpsGeq1}
\end{minipage}
\end{figure}
\end{center}
Note that for the fourth figure, the general form of the congruence is the same for $\varepsilon >1$ and for $\varepsilon =1$, independently of the integrability of $M(u)$ near $u=+\infty$, because all curves end up at future timelike infinity, whether the limit of $r$ along the curve is positive and finite or infinite.

\end{document}